\documentclass[aps,prl,reprint,amsmath,amssymb,superscriptaddress,showpacs]{revtex4-2}
\usepackage{graphicx}
\usepackage{dcolumn}
\usepackage{bm}
\usepackage[colorlinks,linkcolor=blue,anchorcolor=blue,citecolor=blue,urlcolor=blue,filecolor=blue,menucolor=blue,runcolor=blue]{hyperref}

\begin{document}
\title{ Quasiparticle characteristics of the weakly ferromagnetic Hund's metal MnSi }
\author{Yuan Fang}
\thanks{These authors contributed equally to this paper}
\affiliation{Center for Correlated Matter and Department of Physics, Zhejiang University, Hangzhou 310058, China}
\author{Huali Zhang}
\thanks{These authors contributed equally to this paper}
\affiliation{Center for Correlated Matter and Department of Physics, Zhejiang University, Hangzhou 310058, China}
\author{Ding Wang}
\thanks{These authors contributed equally to this paper}
\affiliation{Center for Correlated Matter and Department of Physics, Zhejiang University, Hangzhou 310058, China}
\author{Guowei Yang}
\affiliation{Center for Correlated Matter and Department of Physics, Zhejiang University, Hangzhou 310058, China}
\author{Yi Wu}
\affiliation{Center for Correlated Matter and Department of Physics, Zhejiang University, Hangzhou 310058, China}
\author{Peng Li}
\affiliation{Center for Correlated Matter and Department of Physics, Zhejiang University, Hangzhou 310058, China}
\author{Zhiguang Xiao}
\affiliation{Center for Correlated Matter and Department of Physics, Zhejiang University, Hangzhou 310058, China}
\author{Tianyun Lin}
\affiliation{Center for Correlated Matter and Department of Physics, Zhejiang University, Hangzhou 310058, China}
\author{Hao Zheng}
\affiliation{Center for Correlated Matter and Department of Physics, Zhejiang University, Hangzhou 310058, China}
\author{Xiao-Long Li}
\affiliation{Shanghai Synchrotron Radiation Facility, Shanghai Advanced Research Institute, Chinese Academy of Sciences, Shanghai 201204, China}
\author{Huan-Hua Wang}
\affiliation{Institute of High Energy Physics, Chinese Academy of Sciences, Beijing 100049, China}
\author{Fanny Rodolakis}
\affiliation{Advanced Photon Source, Argonne National Laboratory, 9700 South Cass Avenue, Argonne, Illinois 60439, USA}
\author{Yu Song}
\affiliation{Center for Correlated Matter and Department of Physics, Zhejiang University, Hangzhou 310058, China}
\author{Yilin Wang}
\affiliation{Hefei National Laboratory for Physical Sciences at Microscale, University of Science and Technology of China, Hefei, Anhui 230026, China}
\author{Chao Cao}
\affiliation{Center for Correlated Matter and Department of Physics, Zhejiang University, Hangzhou 310058, China}
\author{Yang Liu}
\email {yangliuphys@zju.edu.cn}
\affiliation{Center for Correlated Matter and Department of Physics, Zhejiang University, Hangzhou 310058, China}
\affiliation{Zhejiang Province Key Laboratory of Quantum Technology and Device, Zhejiang University, Hangzhou, China}
\affiliation{Collaborative Innovation Center of Advanced Microstructures, Nanjing University, Nanjing 210093, China}
\date{\today}%
\addcontentsline{toc}{chapter}{Abstract}

\begin{abstract}
  Hund's metals are multi-orbital systems with $3d$ or $4d$ electrons exhibiting both itinerant character and local moments, and they feature Kondo-like screenings of local orbital and spin moments, with suppressed coherence temperatures driven by Hund's coupling $J_H$. They often exhibit magnetic order at low temperature, but how the interaction between the Kondo-like screening and long-range magnetic order is manifested in the quasiparticle spectrum remains an open question. Here we present spectroscopic signature of such interaction in a Hund's metal candidate MnSi exhibiting weak ferromagnetism. Our photoemission measurements reveal renormalized quasiparticle bands near the Fermi level with strong momentum dependence: the ferromagnetism manifests through possibly exchange-split bands (Q1) below $T_C$, while the spin/orbital screenings lead to gradual development of quasiparticles (Q2) upon cooling. Our results demonstrate how the characteristic spin/orbital coherence in a Hund's metal could coexist and compete with the magnetic order to form a weak itinerant ferromagnet, via quasiparticle bands that are well separated in momentum space and exhibit distinct temperature dependence. Our results imply that the competition between the spin/orbital screening and the magnetic order in a Hund's metal bears interesting similarities to the Kondo lattice systems.
\end{abstract}

\maketitle

Understanding weak itinerant ferromangets represents an outstanding question in condensed matter physics \cite{Santiago2017itinerantmm,brando2016metallic}. In these materials, the ordered moments in the ferromagnetic (FM) phase are much smaller than the local moments inferred from the Curie-Weiss law above the Curie temperature $T_C$, which indicates that models based on the local-moment physics (e.g. the Heisenberg model) do not hold. On the other hand, a simple itinerant band picture is also inadequate, as it does not capture effects due to spin fluctuations and electronic correlations. As an archetypal weak ferromagnet, MnSi plays an important role in the understanding of weak ferromagnetism and validating the famous spin fluctuation theory by Moriya \cite{Moriya1985spinfluctuation}. Nevertheless, the importance of electron correlation and the underlying mechanism of coexisting local moments and itinerant electrons remain open questions.

In multi-orbital ($3d$ or $4d$) weak ferromagnets, strong electron correlations could arise from the inter-orbital Hund's coupling $J_H$, leading to the so-called Hund's metal (HM) \cite{Yin2011kinetic,Yin2011magnetism,Medici2011janus,Mravlje2011coherence,Georges2013hundcoupling}. The HM physics is thought to play an important role in many correlated electron systems, including iron pnictides, ruthenates, transition metal chalcogenides, etc \cite{Yin2011kinetic,Yin2011magnetism,Medici2011janus,Mravlje2011coherence,Georges2013hundcoupling,Yi2013Observation,Miao2016Orbital,Damascelli2000Fermi,Wang2004Quasiparticle,Kondo2016Orbital,Tamai2019high,Jang2021direct}. A recent study from inelastic neutron scattering (INS) and dynamic mean-field theory (DMFT) suggested that MnSi is a HM exhibiting strong orbital and spin fluctuations \cite{chen2020unconvhundmetal}, which can be crucial for its non-Fermi-liquid behaviors over a large phase space \cite{Pfleiderer2001NFL,Doiron2003FLbreak,Pfleiderer2004partialorder,Ritz2013nFL}. It was further proposed that the orbital coherence scale might be smaller than the spin one in MnSi, implying possibly an unconventional type of HM as a result of strong electron correlations \cite{chen2020unconvhundmetal}. A key characteristic of the HM is the quasiparticle (QP) bands that emerge gradually at low temperature, due to the low coherence temperature, but these QP bands remain elusive in MnSi. In addition, how the QP coherence as a result of the HM physics interacts with the weak FM is another interesting question. While a previous study from angle-resolved photoemission spectroscopy (ARPES) on bulk MnSi(001) suggested weak electron correlation \cite{Kura2008MnSi100}, another ARPES study on thick MnSi(111) films emphasized the importance of Fermi surface (FS) nesting in driving the strong magnetic fluctuations \cite{Nicolaou2015arpes}. Nevertheless, neither the FM exchange splitting below $T_C$ nor the characteristic QP bands as a result of the HM physics have been identified so far.

\begin{figure}[ht]
\includegraphics[width=1.\columnwidth]{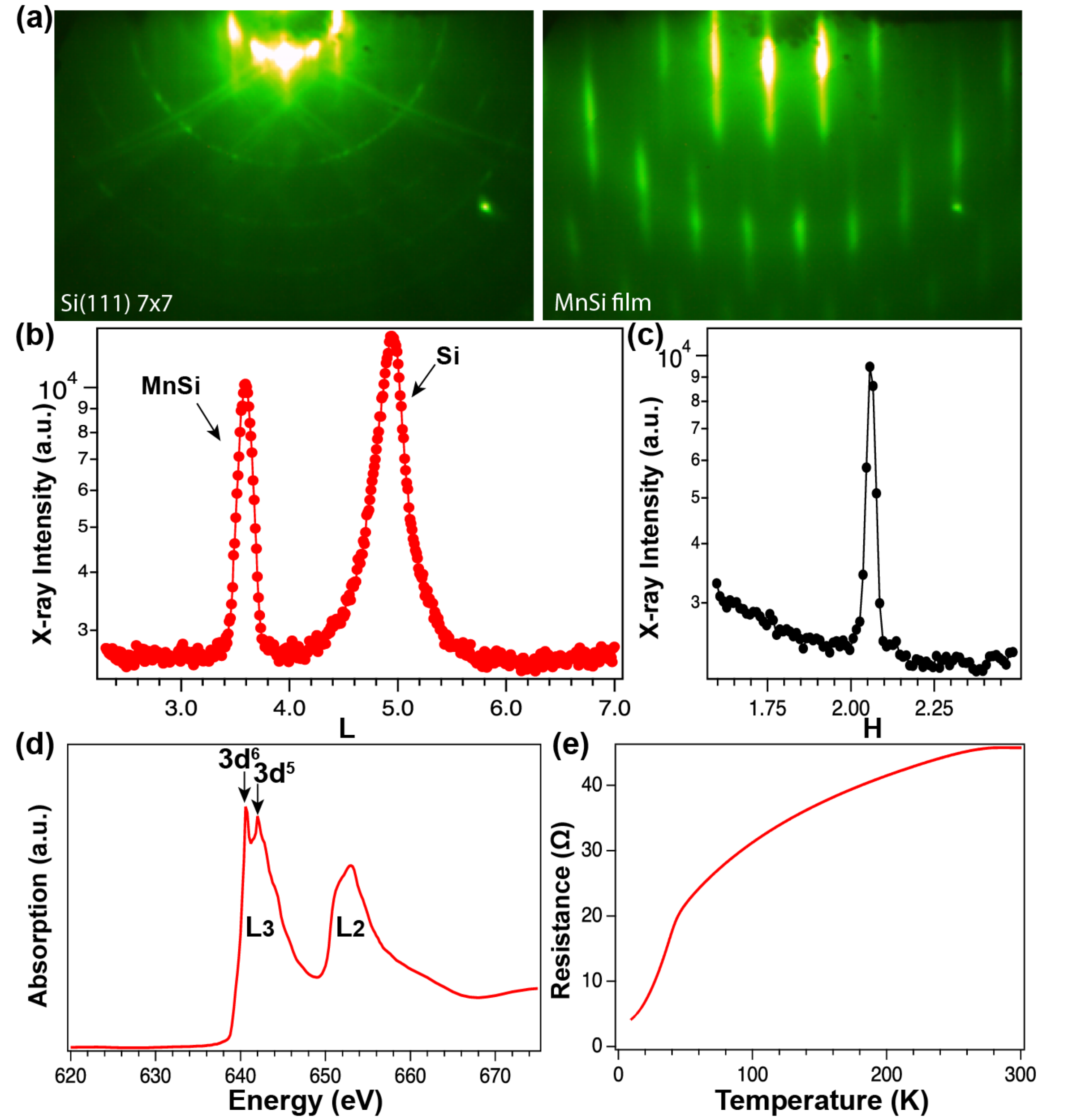}
\centering
\caption{Characterization of MnSi films. (a) RHEED patterns of the Si(111) 7x7 surface and a typical MnSi film. (b,c) X-ray diffraction scans of a thick MnSi film near the MnSi Bragg peak centered at (H,K,L)=(2.069,0,3.582), where H,K,L are defined in units of Si substrate \cite{supplementary}. (d) XAS spectrum near the Mn $L$ edge taken at 29 K. (e) Resistance vs temperature for a MnSi film, showing a kink at $T_C$$\sim$45 K.
}
\label{fig1}
\end{figure}

Here, combining thin film growth by molecular beam epitaxy (MBE) and $in$-$situ$ electronic structure measurements by ARPES, we demonstrate the QP characteristics of the weakly ferromagnetic HM MnSi. Our MnSi films were grown on Si(111) substrates and transferred under ultrahigh vacuum to a Helium-lamp ARPES system for electronic structure measurements. Details can be found in the Supplemental Material \cite{supplementary} (see also references \cite{Geisler2012Growth,PhysRevB.47.558,PhysRevB.59.1758} therein). Figure 1(a) shows the reflection high energy electron diffraction (RHEED) patterns from the starting Si(111)-7x7 surface and a as-grown thick MnSi film, confirming epitaxial growth of high-quality MnSi films. The epitaxial growth is further confirmed by $ex$-$situ$ X-ray scattering measurements (Fig. 1(b,c)), showing that the in-plane and out-of-plane lattice constants are $\sim$3.243 {\AA} and $\sim$7.872 {\AA}, respectively \cite{supplementary}. This indicates that the epitaxial MnSi film undergoes a tensile strain of $\sim$1$\%$ compared to bulk MnSi, although it does not fully match with the underlying Si substrate, similar to previous studies \cite{Nicolaou2015arpes,Karhu2010Structure}. The strain gives rise to an enhanced $T_C$$\sim$45 K (compared to $T_C$$\sim$30 K in bulk MnSi), as evidenced by the characteristic kink in the resistivity vs temperature data shown in Fig. 1(e). Due to the non-centrosymmetric crystal structure, there is a weak Dzyaloshinskii-Moriya (DM) interaction which leads to a very small rotation ($\sim$5$^{\circ}$) of the in-plane Mn moments between the adjacent (111) planes \cite{Shirane1983spiral}. However, the DM interaction is too weak to generate any obvious effect on the observed QP bands (see Fig. S1 in \cite{supplementary}), and therefore is ignored in this paper. Measurements from X-ray absorption spectroscopy (XAS) near the Mn $L$ edge indicate that the ground state is mixed-valent with both $3d^5$ and $3d^6$ configurations, similar to bulk MnSi \cite{Carbone2006eeinteraction}. The presence of the $3d^6$ configuration is unusual for Mn and is likely caused by hybridization with Si orbitals. Note that the $3d^6$ configuration is one electron more than half filling, implying an important role of Hund's physics \cite{chen2020unconvhundmetal}.

\begin{figure*}[ht]
\centering
\includegraphics[width=2.\columnwidth]{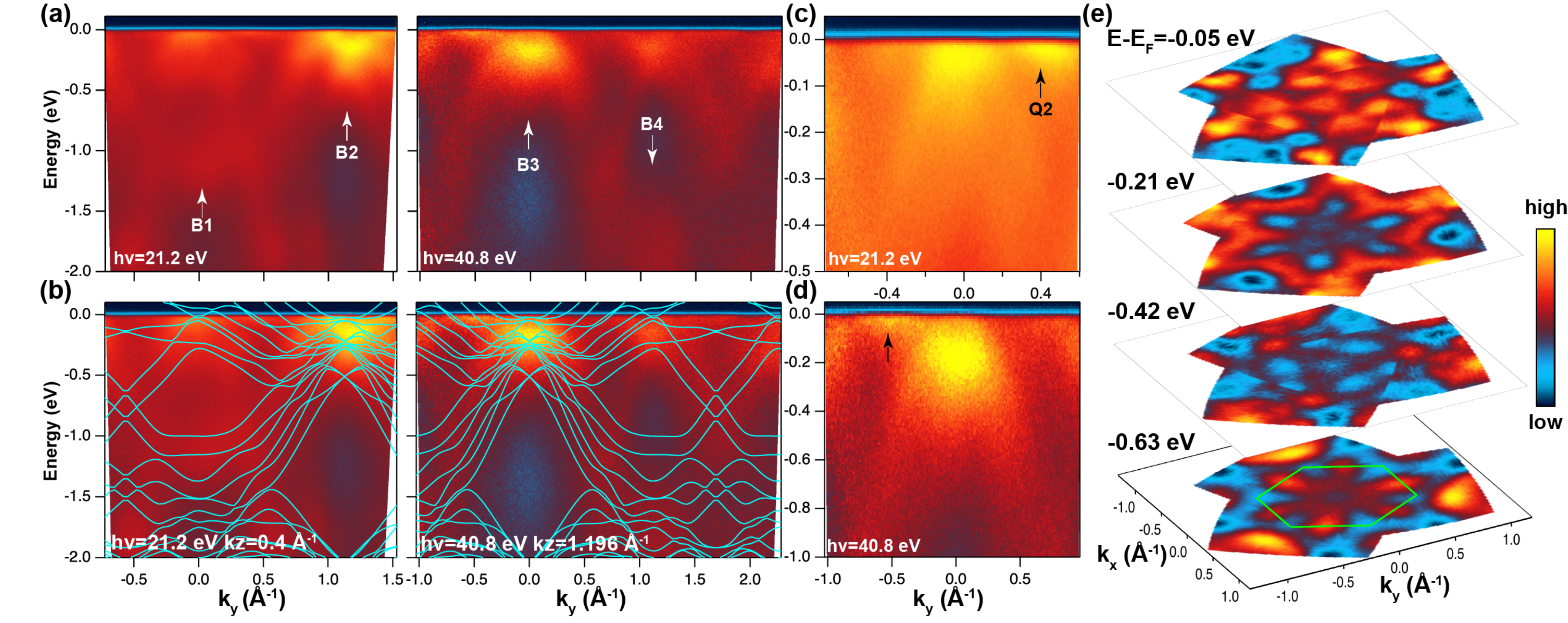}
\caption{ARPES spectra of MnSi films taken at $\sim$6 K. (a) ARPES data of a MnSi film taken with 21.2 eV and 40.8 eV photons, measured along $\Bar{\varGamma}$-$\bar{M}$. (b) Results from DFT calculations overlaid on top of ARPES data. The $k_z$ values are based on an estimated inner potential of $\sim$12.8 eV. (c,d) Zoomed-in view of the QP bands near $E_F$. Black arrows highlight flat QP bands near $E_F$. (e) Constant energy maps at various energies using 21.2 eV photons. The green hexagon at the bottom marks the surface BZ.
}
\label{fig2}
\end{figure*}

Figure 2(a) shows the energy-momentum dispersion taken with He I (21.2 eV) and He II (40.8 eV) photons, revealing dispersive QP bands near the Fermi level ($E_F$). For comparison, Fig. 2(b) shows the calculated band structure from density functional theory (DFT) overlaid on top of the ARPES data, with $k_z$ = 0.4 {\AA}$^{-1}$ and 1.196 {\AA}$^{-1}$ based on an estimated inner potential of $\sim$12.8 eV {(see Fig. S2 in \cite{supplementary} for more details)}. The Mn moment was set to the experimental value of 0.4 $\mu_B$/Mn in DFT calculations \cite{Jeong2004MnSiB20}, which yields a FM exchange splitting of $\sim$0.2 eV (see Fig. S3 in \cite{supplementary}). The overall valence bands {from the 21.2 eV data} can be partially explained by the DFT calculations, particularly the broad parabolic band near -1.2 eV {(labelled B1)} centered at $\Bar{\varGamma}$ and the hole-like bands near $E_F$ {(labelled B2)} centered at $k_y\sim$1.12 {\AA}$^{-1}$, i.e., $\Bar{\varGamma}$ in the second surface Brillouin zone (BZ). {For the 40.8 eV data, different set of bands (labelled B3,B4 in Fig. 2(a)) can be observed}. The obvious spectral difference between the first and second surface BZs confirms that the observed spectral features are bulk states, as the bulk BZ is twice as large as the surface BZ (see Fig. 2(b)). Despite these similarities between ARPES results and DFT calculations, the ARPES spectra also show well-defined flat bands right near $E_F$ extending over a large momentum region ($\pm$0.6 {\AA}$^{-1}$), {marked by black arrows} in Fig. 2(c,d), which cannot be easily accounted for by DFT calculations. {We mention that DMFT calculations indeed suggest enhanced effective mass (or flatness) of quasiparticle bands near $E_F$ compared to DFT \cite{chen2020unconvhundmetal}, although they cannot reproduce quantitatively the pronounced flat bands observed experimentally (note that DMFT calculations in \cite{chen2020unconvhundmetal} were performed at room temperature).} In addition, the experimental constant energy maps (Fig. 2(e)) show obvious deviations from the DFT calculations (see Fig. S4 in \cite{supplementary}). Such a discrepancy implies appreciable band renormalization near $E_F$, suggesting that correlation effects are likely important in MnSi. The flat QP bands near $E_F$ are consistent with the enhanced effective mass from de Haas-van Alphen \cite{Tailefer1986Band,Wilde2021symmetry} and specific heat measurements \cite{Bauer2010quantum,Mishra2016low}. Note that the QP bands are well-defined near $E_F$ and become much broader away from $E_F$, in agreement with the strong QP scattering as a result of correlation effect.

\begin{figure*}[ht]
\centering
\includegraphics[width=2.\columnwidth]{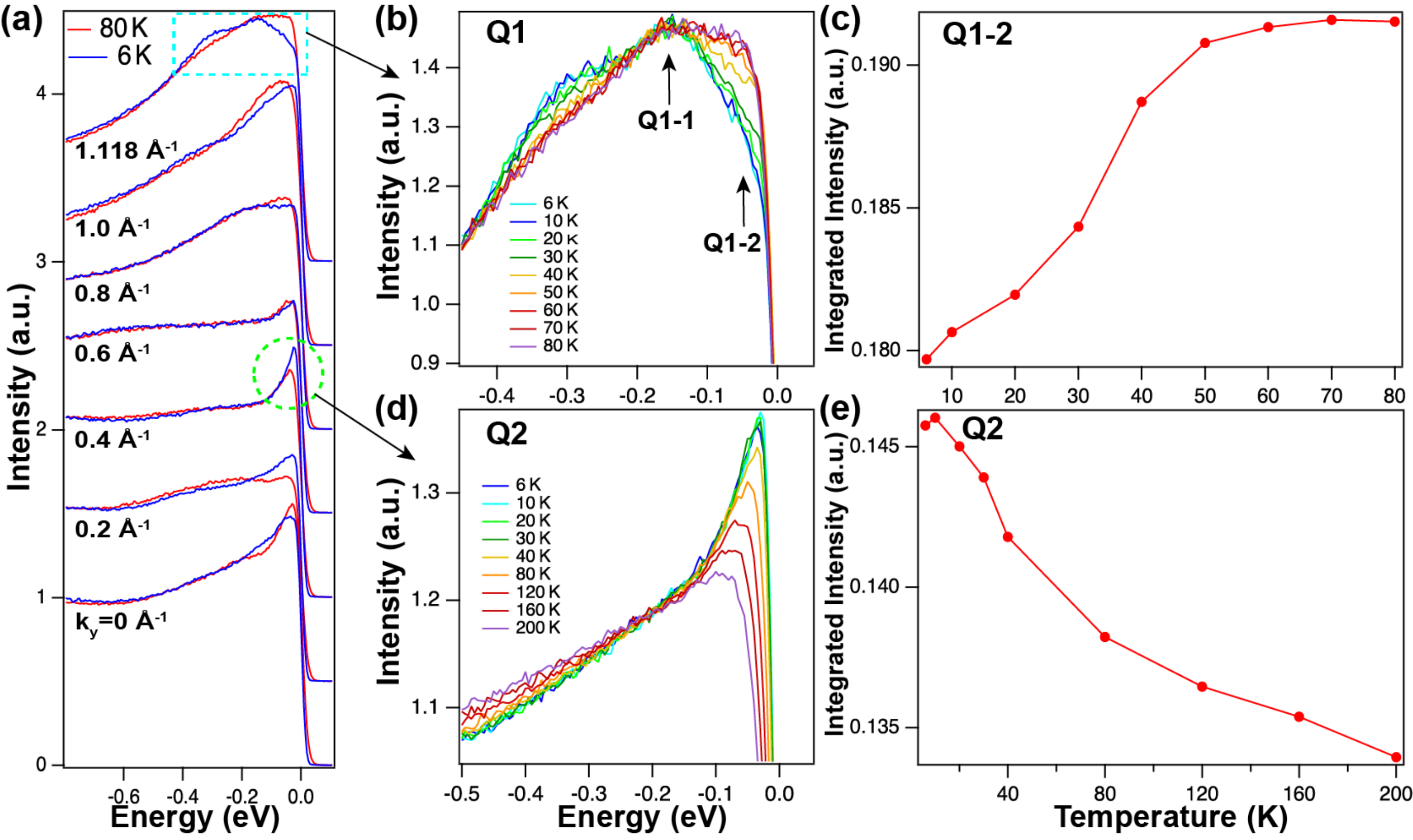}
\caption{Temperature evolution of QP bands. (a) Momentum-dependent EDCs well below (blue curves, $\sim$6 K) and above $T_C$ (red curves, $\sim$80 K) along $\Bar{\varGamma}$-$\bar{M}$ direction. The cyan dashed box and green dashed circle highlight the Q1 (b,c) and Q2 (d,e) bands, respectively. (b,d) Temperature evolution of the EDCs. (c,{e}) The integrated intensity of Q1-2 (c) and Q2 ({e}) from (b,d) as a function of temperature. The integration region is [-0.145 eV, -0.01 eV] for Q1-2 and [-0.125 eV, 0.005 eV] for Q2.
}
\label{fig3}
\end{figure*}

To understand the electron correlation from the HM physics and the origin of FM, the temperature evolution of the QP bands is crucial. Fig. 3(a) shows the momentum-dependent energy distribution curves (EDCs) along the $\Bar{\varGamma}$-$\bar{M}$ direction above and below $T_C$, where clear spectral changes across $T_C$ with strong momentum dependence can be observed. In particular, near $k_y\sim$1.118 {\AA}$^{-1}$ (Fig. 3(b)), there is a broad asymmetric peak (Q1) at $\sim$-0.15 eV in the paramagnetic (PM) phase (see the 80 K data), which evolves into two peaks at $\sim$-0.3 and $\sim$-0.15 eV in the FM phase, and at the same time, the spectral weight at $E_F$ is reduced (see the 6 K data). According to DFT calculations in the PM phase (see Fig. S3(a) in \cite{supplementary}), the broad asymmetric Q1 peak could contain contributions from two close-by bands, i.e., Q1-1 at $\sim$-0.15 eV and Q1-2 very close to $E_F$. Upon entering the FM phase, the FM exchange splittings of Q1-1 and Q1-2 could lead to emergence of a new satellite band at $\sim$-0.3 eV (from the majority band of Q1-1), a more pronounced central peak at $\sim$-0.15 eV (due to overlapping contributions from split Q1-1 and Q1-2 peaks) and reduced density of states (DOS) at $E_F$ (as the minority band of Q1-2 is pushed above $E_F$). See Fig. S6 in \cite{supplementary} for a schematic illustration of our proposed scenario. Detailed temperature-dependent scans (Fig. 3(b,c)) further show that this temperature-driven transition takes place near $T_C$$\sim$45 K, consistent with its FM origin. Although our interpretation of the Q1 temperature evolution in terms of FM exchange splitting is the most plausible one based on current data, other possibilities cannot be completely ruled out. In contrast to a momentum-independent FM exchange splitting of $\sim$0.2 eV obtained from DFT calculations \cite{supplementary}, the experimental FM splitting seems to be highly anisotropic in momentum space. For instance, the QP band near $k_y\sim$ 0.6 {\AA}$^{-1}$ shows almost no sign of FM splitting, {while the EDCs at $k_y\sim$ 0.4 and 0.2 {\AA}$^{-1}$ show the opposite trend compared to that at $k_y\sim$ 1.118 {\AA}$^{-1}$ (Fig. 3(a)), i.e., the spectral weight is now transferred from deeper energies to $E_F$ upon cooling. This leads to development of sharp quasiparticle peak near $E_F$ at low temperature, e.g., the Q2 peak at $k_y\sim$0.4 {\AA}$^{-1}$.}

Although it is tempting to attribute the observed momentum anisotropy to momentum-dependent FM exchange splitting, as observed in some FM systems \cite{Sanchez2012effects,Miyazaki2009EuO,Kim2015sro}, the gradual enhancement of the Q2 peak upon cooling over a wide temperature range implies that its spectral change is not caused by the FM order (Fig. 3(d)). Instead, such behavior is consistent with the QPs expected in a HM, i.e., QPs with low coherence temperature and energy scale. The suppressed coherence temperature is due to Hund's coupling, i.e., large Hund's coupling will help formation of large local moments with low coherence temperature \cite{Georges2013hundcoupling}. Interestingly, the growth of the Q2 peak appears to slow down below $\sim$30 K, as shown in Fig. 3(d). To recover the full spectral function near $E_F$, we divide the temperature-dependent EDCs by the resolution-convoluted Fermi-Dirac function (for details, see Fig. S5 in \cite{supplementary}). The integrated peak intensity after such analysis is summarized in Fig. 3(e), which indicates that the peak intensity grows monotonically above $T_C$, and shows sign of slowing down well below $T_C$ \cite{supplementary}.

It is interesting to note that the Q2 QP development at low temperature is limited to a small energy region near $E_F$ (Fig. 3(d)), and its intensity over a wide temperature range can be roughly described by the $-\log(T)$ dependence (Fig. S5 in \cite{supplementary}). These characteristics bear intriguing similarity to the renowned Kondo resonances in KL systems \cite{denlinger2001comparative,allen2005kondo,fujimori2016band,Kirchner2020Colloqium}. There, the spin-flip scattering of conduction electrons with the local moments leads to spin screening and a gradual buildup of the Kondo resonance near $E_F$. In the HM, the screening could occur in both the spin and orbital channels \cite{Deng2019motthund,chen2020unconvhundmetal}, with separated energy scales, which can give rise to QP bands with strong temperature dependence, as exemplified in ruthenates \cite{Wang2004Quasiparticle,Kondo2016Orbital,Tamai2019high,liu2018heavyfermiliquid}. The low coherence energy also results in a broad regime with bad metal or non-Fermi-liquid (NFL) behaviors \cite{Walter2020nFLhund}. Indeed, MnSi is well-known for its extended NFL region over a large pressure/temperature region \cite{Pfleiderer2001NFL,Doiron2003FLbreak,Pfleiderer2004partialorder,Ritz2013nFL}.

The possible slowdown of the Q2 QP development below $T_C$ (see Fig. S5(i) in \cite{supplementary}) suggests that the QP coherence as a result of orbital/spin screening might be interrupted by the FM order. This implies interesting analogy to the $f$-electron KL systems: there, the magnetic order (via the Ruderman-Kittel-Kasuya-Yosida (RKKY) exchange interaction) competes with the heavy QP formation (via the Kondo screening), resulting in the well-known Doniach phase diagram that lies at the heart of the heavy fermion physics \cite{Doniach1977Kondo}. Often, magnetic KL systems exhibit some degree of Kondo screening, and recent ARPES studies on CeSb and USb$_2$ indicate that the competition between the RKKY and Kondo interactions is manifested by the momentum-separated QP bands \cite{Jang2019Direct,Chen2019Orbital}. The itineracy of the $f$ electrons can be estimated by the magnitude of the screened moment, i.e., the ratio between the local moment above $T_C$ and the ordered moment below $T_C$. In bulk MnSi, this ratio is $\sim$5 \cite{Yasuoka1978NMR,Shirane1983spiral}: such moderately large value is probably a direct consequence of the competition between the spin/orbital screening from the HM physics and the long-range FM order. Our results therefore suggest that the FM order and the spin/orbital screening can coexist and compete in a multi-orbital $3d$ HM, via momentum-separated QP bands with distinct temperature dependence, resulting in a weak itinerant ferromagnet. Such insight might help explain the dichotomy of itinerant and local moments observed in recent INS studies \cite{chen2020unconvhundmetal,Jin2021Magnetic}.

To conclude, we present systematic temperature-dependent ARPES results for a HM candidate MnSi with weak FM. We observed well-defined QP bands with pronounced temperature dependence: the Q1 band shows a possible FM exchange splitting below $T_C$, originating from its weak itinerant FM; by contrast, the Q2 peak grows monotonically upon cooling above $T_C$, and becomes fully developed inside the FM phase, as a result of characteristic orbital/spin screening in a HM. Our results therefore provide spectroscopic insight for understanding the coexistence and competition between the weak FM (forming ordered moments) and the HM physics (favoring screened moments at low temperature): it can be achieved through QP bands that are well separated in momentum space and exhibit distinct temperature dependence. Such competition in a ferromagnetic HM bears striking similarity with the classical KL systems. Our study motivates future works to understand the underlying mechanism of the momentum-dependent electron correlation, as well as the inherent connection between the Kondo physics and HM physics. Interestingly, a very recent DMFT study shows that the QP self-energy can indeed exhibit strong momentum anisotropy in a multi-orbital Hund's ferromagnet \cite{Nomura2022Fermi}.

\section{acknowledgments}
 This work is supported by the Key R$\&$D Program of Zhejiang Province, China (2021C01002), the State Key project of Zhejiang Province (No. LZ22A040007), the National Science Foundation of China (No. 12174331, 11674280), the National Key R$\&$D Program of China (Grant No. 2017YFA0303100, 2016YFA0300203) and the Fundamental Research Funds for the Central Universities (2021FZZX001-03). The surface X-ray diffraction experiments were performed at Beamline 1W1A in Beijing Synchrotron Radiation Facility (BSRF) and Beamline BL02U2 in Shanghai Synchrotron Radiation Facility (SSRF). The XAS measurements were performed at 29-ID IEX beamline at Advanced Photon Source (APS). APS is supported by the US Department of Energy (DOE) Office of Science under Contract No. DE-AC02-06CH11357; additional support by the National Science Foundation under Grant No. DMR-0703406 is also acknowledged. We thank Dr. Chenchao Xu, Prof. Xin Lu, Prof. Huiqiu Yuan, Prof. Zhi-Cheng Zhong, Prof. Yi-feng Yang for experimental help and useful discussions.

\end{document}


\title{ Supplementary Information for: \\ Quasiparticle characteristics of the weakly ferromagnetic Hund's metal MnSi }
\author{Yuan Fang}
\thanks{These authors contributed equally to this paper}
\affiliation{Center for Correlated Matter and Department of Physics, Zhejiang University, Hangzhou 310058, China}
\author{Huali Zhang}
\thanks{These authors contributed equally to this paper}
\affiliation{Center for Correlated Matter and Department of Physics, Zhejiang University, Hangzhou 310058, China}
\author{Ding Wang}
\thanks{These authors contributed equally to this paper}
\affiliation{Center for Correlated Matter and Department of Physics, Zhejiang University, Hangzhou 310058, China}
\author{Guowei Yang}
\affiliation{Center for Correlated Matter and Department of Physics, Zhejiang University, Hangzhou 310058, China}
\author{Yi Wu}
\affiliation{Center for Correlated Matter and Department of Physics, Zhejiang University, Hangzhou 310058, China}
\author{Peng Li}
\affiliation{Center for Correlated Matter and Department of Physics, Zhejiang University, Hangzhou 310058, China}
\author{Zhiguang Xiao}
\affiliation{Center for Correlated Matter and Department of Physics, Zhejiang University, Hangzhou 310058, China}
\author{Tianyun Lin}
\affiliation{Center for Correlated Matter and Department of Physics, Zhejiang University, Hangzhou 310058, China}
\author{Hao Zheng}
\affiliation{Center for Correlated Matter and Department of Physics, Zhejiang University, Hangzhou 310058, China}
\author{Xiao-Long Li}
\affiliation{Shanghai Synchrotron Radiation Facility, Shanghai Advanced Research Institute, Chinese Academy of Sciences, Shanghai 201204, China}
\author{Huan-Hua Wang}
\affiliation{Institute of High Energy Physics, Chinese Academy of Sciences, Beijing 100049, China}
\author{Fanny Rodolakis}
\affiliation{Advanced Photon Source, Argonne National Laboratory, 9700 South Cass Avenue, Argonne, Illinois 60439, USA}
\author{Yu Song}
\affiliation{Center for Correlated Matter and Department of Physics, Zhejiang University, Hangzhou 310058, China}
\author{Yilin Wang}
\affiliation{Hefei National Laboratory for Physical Sciences at Microscale, University of Science and Technology of China, Hefei, Anhui 230026, China}
\author{Chao Cao}
\affiliation{Center for Correlated Matter and Department of Physics, Zhejiang University, Hangzhou 310058, China}
\author{Yang Liu}
\email {yangliuphys@zju.edu.cn}
\affiliation{Center for Correlated Matter and Department of Physics, Zhejiang University, Hangzhou 310058, China}
\affiliation{Zhejiang Province Key Laboratory of Quantum Technology and Device, Zhejiang University, Hangzhou, China}
\affiliation{Collaborative Innovation Center of Advanced Microstructures, Nanjing University, Nanjing 210093, China}
\date{\today}%
\maketitle
\section{MBE thin film growth and $IN$-$SITU$ ARPES measurement}
Our growth method of epitaxial MnSi films was adapted from previous studies [\textcolor{blue}{22,24,27}], with small adjustments to yield best quality films for ARPES measurements. MnSi films were grown on Si(111) substrates by MBE under ultrahigh vacuum with a pressure better than $2\times10^{-10}$ mbar. The Si(111) substrates were cut into a typical size of 5x10 mm$^2$. The substrates were heated in ultrahigh vacuum to ${900-1000}^{\circ}\mathrm{C}$ to yield the 7x7 reconstructed surface, as verified by RHEED. After that, $\sim$4.2 {\AA} of Mn was deposited on the 7x7 surface at ${70}^{\circ}\mathrm{C}$ or ${200}^{\circ}\mathrm{C}$ and annealed to $\sim$${260}^{\circ}\mathrm{C}$ to form a good buffer layer for subsequent MnSi film growth. We found that a lower growth temperature (${70}^{\circ}\mathrm{C}$) of the buffer layer yields slightly better film quality. Mn was then deposited on the buffer layer at ${260}^{\circ}\mathrm{C}$, allowing Si from the underlying substrate to diffuse to the top layers to form epitaxial MnSi fimls. Codeposition of Mn and Si (from a separate Si effusion cell) was also performed and the ARPES results were found to be similar, despite slightly weaker quasiparticle bands for very thin films. The structural quality of the MnSi films were monitored $in$-$situ$ by RHEED. The deposition of Mn was done with an effusion cell set to ${790}^{\circ}\mathrm{C}$ to yield a rate of approximately 0.7 ML per min, as determined by a quartz crystal monitor. All the data presented in this paper were taken from films with thicknesses ranging from a few nm to a few tens of nm.

All the ARPES data were taken in a Helium-lamp ARPES system at Zhejiang University, which is connected under ultrahigh vacuum to the MBE system. ARPES measurements were performed immediately after the film growth, by transferring the sample from the MBE growth chamber to the ARPES chamber. A five-axis manipulator cooled by a closed-cycle helium refrigerator was employed for sample alignment and for the temperature control during the ARPES measurements. The photon source was a VUV-5k Helium lamp coupled to a grating monochromator for selecting either the He-I (21.2 eV) or He-II (40.8 eV) lines. The base pressure of the ARPES system was $\sim7.6\times10^{-11}$ mbar, which increased to $\sim2.5\times10^{-10}$ mbar during the Helium lamp operation. Most of the ARPES data were taken with He-I photons, unless mentioned otherwise. The typical energy and momentum resolution is $\sim$12 meV and $\sim$0.01 {\AA}$^{-1}$, respectively.

\section{DFT calculations}
Electronic structure calculations were performed using DFT and a plane-wave basis projected augmented wave method, as implemented in the Vienna ab initio simulation package (VASP) [\textcolor{blue}{25,26}]. The Perdew, Burke and Ernzerhof (PBE) generalized gradient approximation for exchange correlation potential was used for the DFT calculation. Spin-orbital coupling (SOC) was taken into account in all DFT calculations. An energy cutoff of 450 eV and a 12×12×12 gamma-centered k-mesh were employed to converge the calculation to 1 meV/atom. The optimized ordered moment for the FM phase of MnSi is 1 $\mu_B$/Mn from the DFT calculation, which is substantially larger than the experimental value of 0.4 $\mu_B$/Mn. For better comparison with experiments, the magnetic moment is constrained to the experimental value, which yields a FM exchange splitting of $\sim$0.2 eV.

\section{X-ray diffraction, transport and X-ray absorption measurements}
The lattice constant of the epitaxial MnSi films were determined by synchrotron X-ray diffraction. The measurements were performed at room temperature for a thick film, in both Beamline 1W1A in Beijing Synchrotron Radiation Facility (BSRF) and Beamline BL02U2 in Shanghai Synchrotron Radiation Facility (SSRF). We used the standard surface coordinate of the substrate Si to present our diffraction data (Fig. 1(b,c)): $a=b=3.84$ {\AA}, $c=9.406$ {\AA}, $\alpha$=$\beta$=${90}^{\circ}$, $\gamma$=${120}^{\circ}$. Therefore, H,K directions lie within the (111) plane, while the L direction points to the out-of-normal [111] direction. From the measured positions of MnSi Bragg peaks, the lattice constants of MnSi film is determined to be $a=b=3.243$ {\AA}, $c=7.872$ {\AA}, $\alpha$=$\beta$=${90}^{\circ}$, $\gamma$=${120}^{\circ}$. This indicates a tensile strain of $\sim$1$\%$ compared with the bulk MnSi.

The resistance measurements were conducted via the standard four-probe method using a quantum design physical property measurement system (PPMS).

The XAS measurements near the Mn $L$ edge were performed using the total electron yield mode at $\sim$29 K, at sector 29 in Advanced Photon Source (APS). A 100 nm MnSi film was employed for the XAS measurements.

\begin{figure}[ht]
\includegraphics[width=.6\columnwidth]{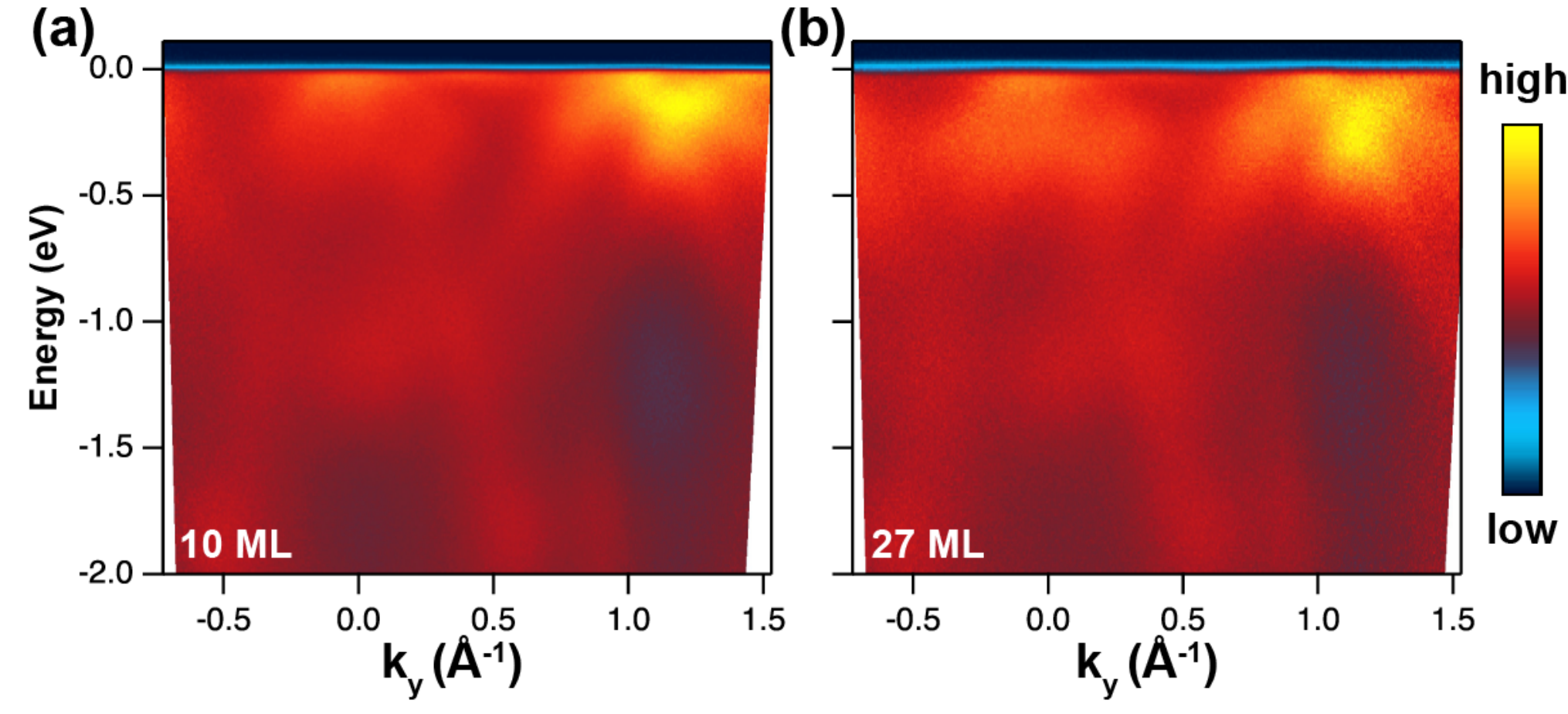}
\caption{ Comparison of ARPES spectra of a 10 monolayer (ML) MnSi film (a) and a 27 ML MnSi film (b). Here one ML (=2.624 {\AA}) refers to the basic repeating structure along the [111] direction, consisting of two Mn and two Si layers with different atomic densities. Note that the helical order has a long periodicity of $\sim$20 nm along the [111] direction, due to very weak DM interaction. Therefore, the film thickness studied here is much smaller than the periodicity of the helix, i.e., the film is not thick enough to accommodate one complete helix. The fact that the ARPES spectra show no obvious change with thickness indicates that the DM interaction plays a minor role on the observed electronic structure.
}
\label {Fig.S1}
\end{figure}

\begin{figure}[ht]
\includegraphics[width=.6\columnwidth]{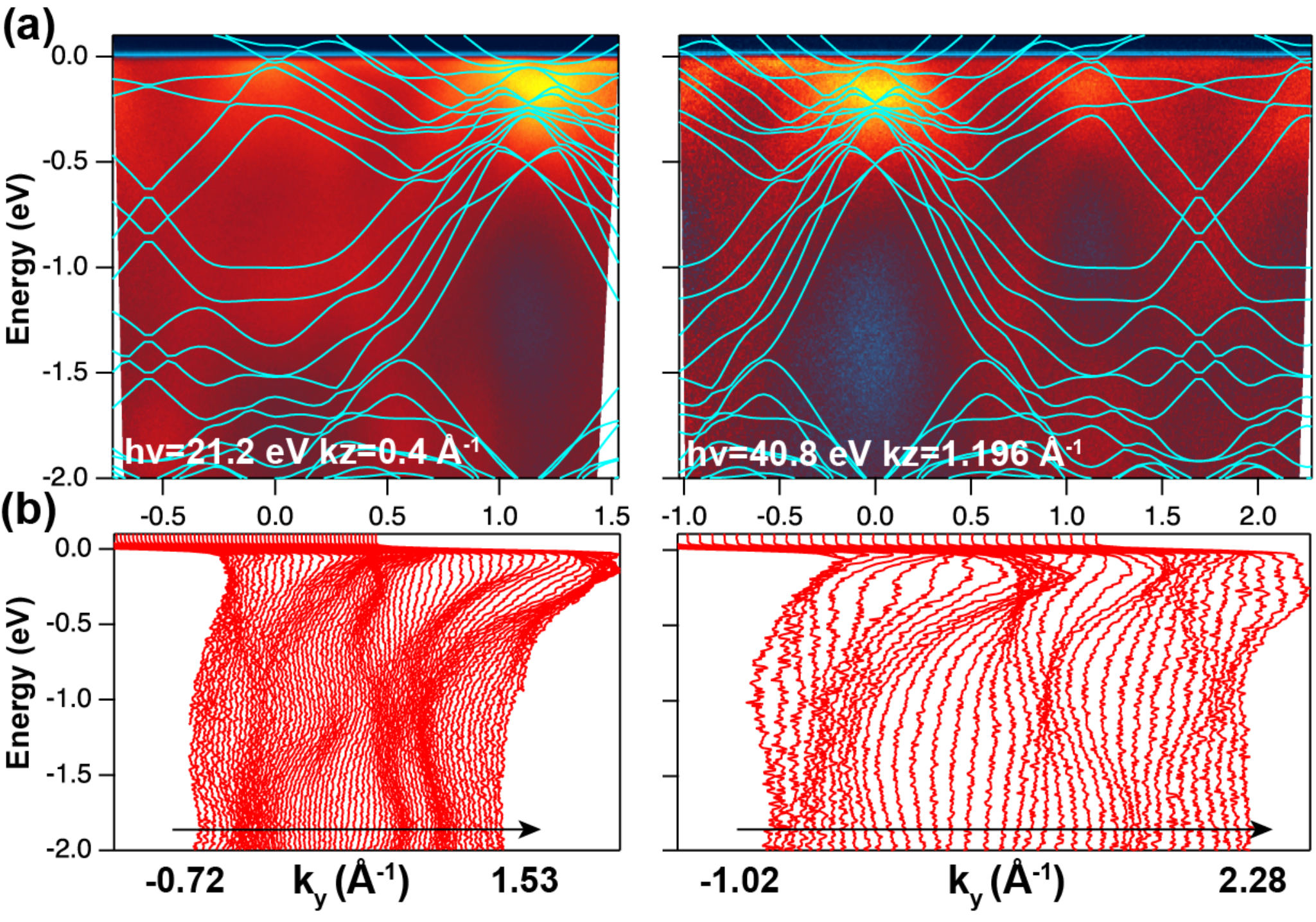}
\caption{ {(a) Direct comparison of the ARPES data and DFT calculations by overlaying the DFT calculations (cyan curves) on top of the experimental data (same as Fig. 2(b)). (b) The corresponding waterfall plot of the EDCs from the ARPES data.
}
}
\label {Fig.S2}
\end{figure}

\begin{figure}[ht]
\includegraphics[width=.8\columnwidth]{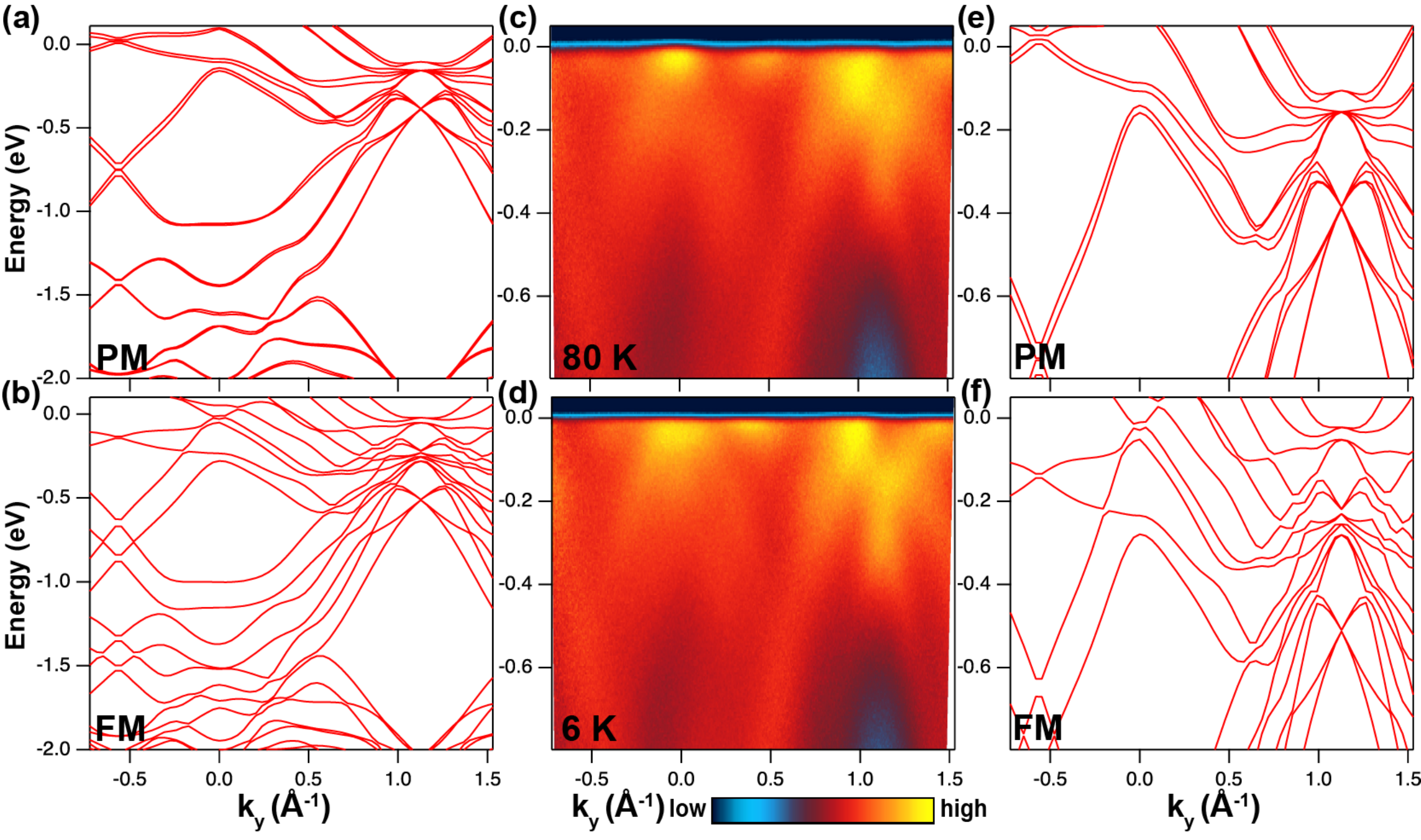}
\caption{ (a,b) Comparison of calculated band structure in a wide energy range for $k_z$ = 0.4 {\AA}$^{-1}$ (corresponding to the $k_z$ cut using He I photons) for the PM (a) and FM (b) phase. A momentum-independent FM exchange splitting of $\sim$0.2 eV can be readily identified from the comparison. Note that the calculated FM exchange splitting is roughly uniform in momentum space, in sharp contrast to the large momentum anisotropy seen in experiments. (c-d) The ARPES spectra taken with He I photons in the PM (c) and FM (d) phases. (e-f) The corresponding DFT calculations in the PM (e) and FM (f) phases for direct comparison with the experimental data in (c,d). The spectral change across $T_C$ is rather small and strongly dependent on the momentum (see EDCs in Fig. 3). The flat QP bands near $E_F$ in the momentum range of $\pm$0.6 {\AA}$^{-1}$ cannot be explained by the DFT calculation.
}
\label {Fig.S3}
\end{figure}

\begin{figure}[ht]
\includegraphics[width=.8\columnwidth]{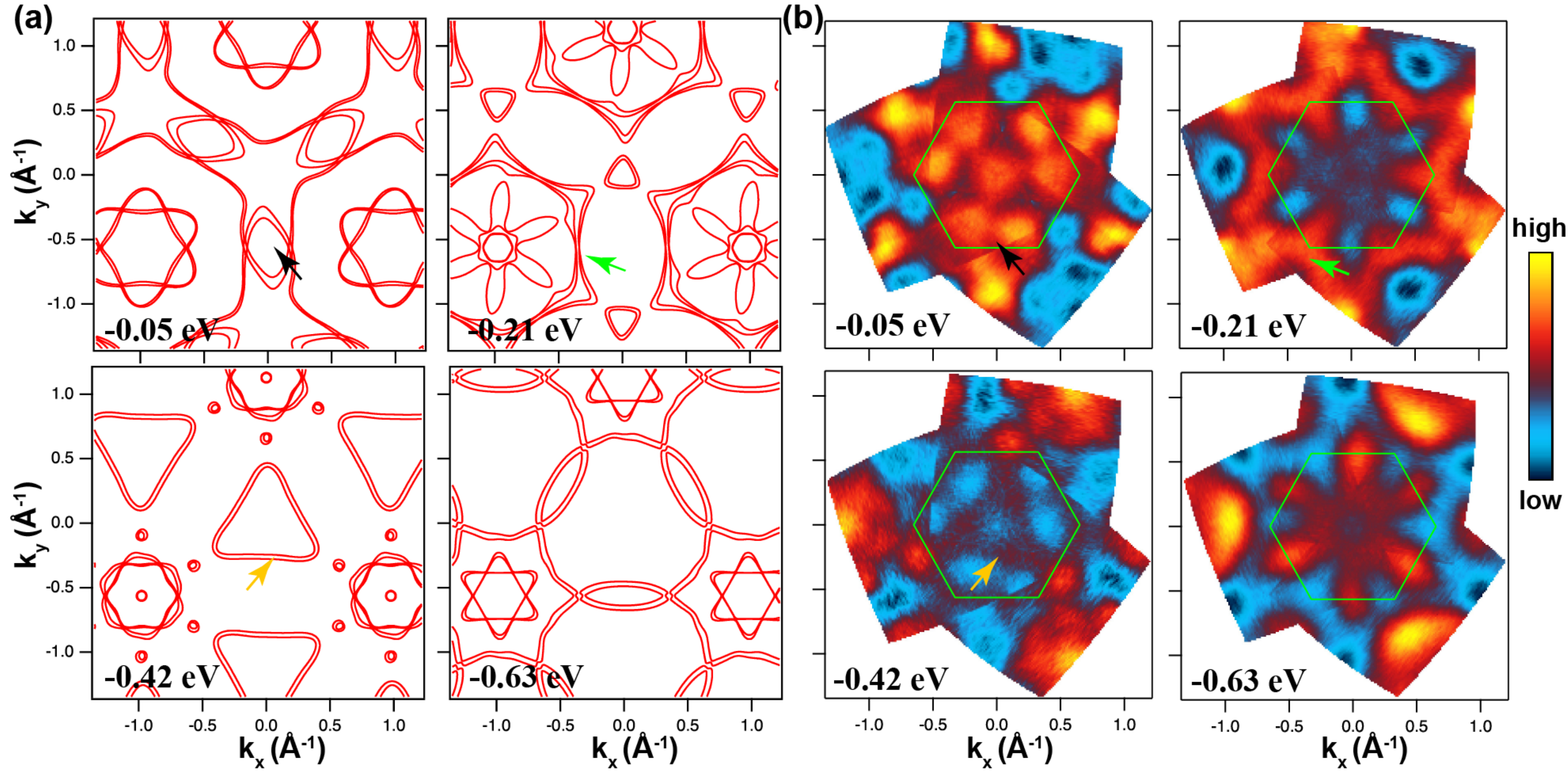}
\caption{ (a) Calculated constant energy contours for MnSi from DFT, for comparison with the experimental results (b) (same as Fig. 2(e)). The green hexagons in (b) indicate surface BZs. {At -0.05 eV, the DFT calculation predicts elliptical pockets centered at $\bar{M}$ (see black arrows), which qualitatively agrees with the ARPES data, although the detailed shapes are quite different. At -0.21 eV, the DFT calculation shows large pockets centered at $\Bar{\varGamma}$ in the second surface Brillouin zone (see green arrows), which is quite different from the experimental data. At -0.42 eV, the DFT calculation predicts a pocket at $\Bar{\varGamma}$ with a triangular shape, which agrees reasonably with the ARPES data. The large difference between DFT and ARPES could be attributed to strong electron correlation and complex band structure of MnSi. }
}
\label {Fig.S4}
\end{figure}

\begin{figure}[ht]
\includegraphics[width=.8\columnwidth]{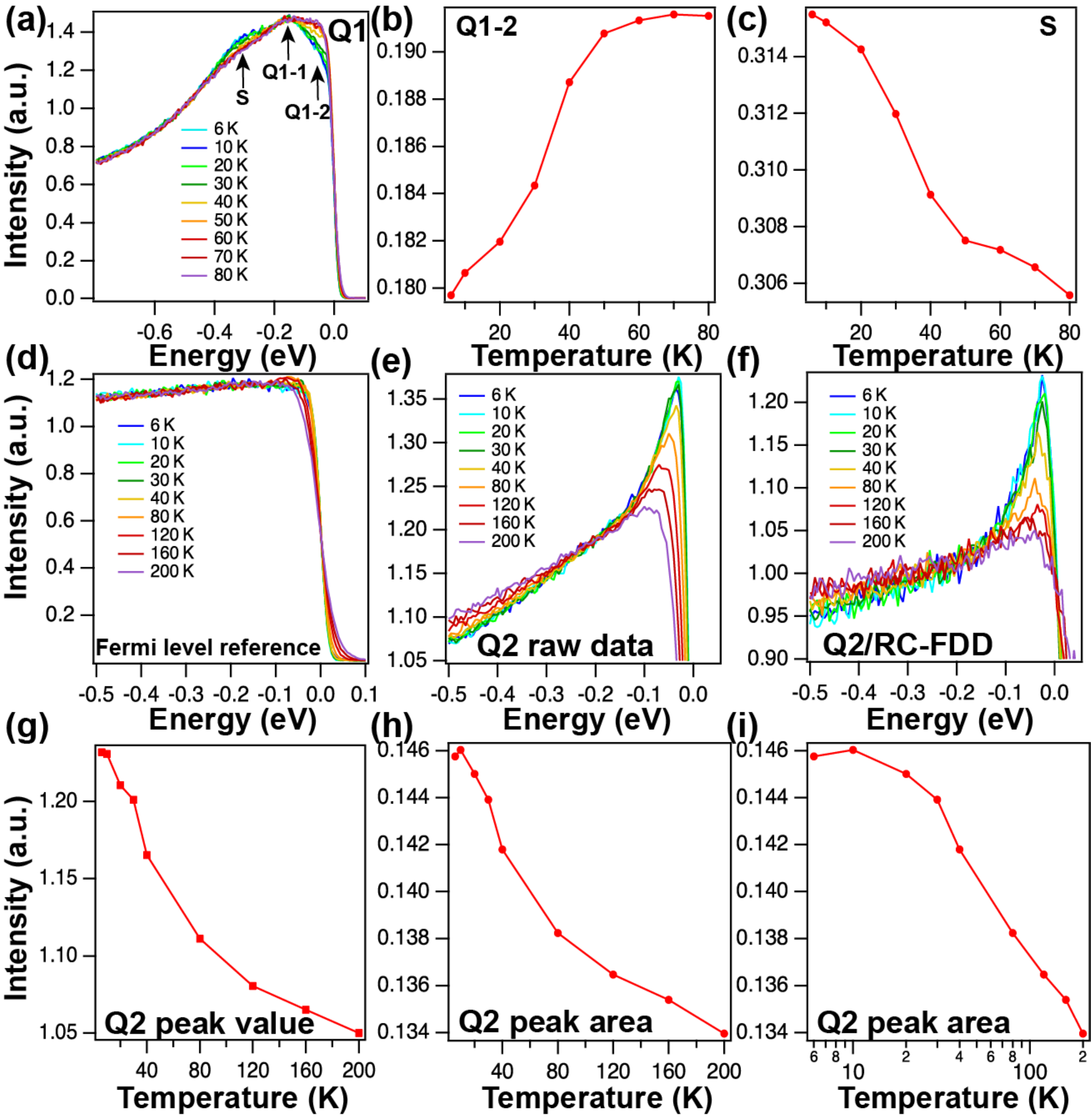}
\caption{ Detailed analysis of the temperature-dependent ARPES data shown in Fig. 3. (a) Temperature-dependent EDCs corresponding to Q1 (same as Fig. 3(b)). {Three spectral features (Q1-1, Q1-2, S) are labelled.} (b,c) Integrated spectral intensity {for Q1-2 and S} marked in (a) as a function of temperature. {The integration range for Q1-2 and S is [-0.145 eV, -0.01 eV] (b) and [-0.445 eV, -0.205 eV] (c), respectively.} (d) The reference spectra for obtaining the resolution-convoluted Fermi-Dirac function (RC-FDD). (e,f) The temperature-dependent EDCs corresponding to the Q2 peak (e) and those divided by the resolution-convoluted Fermi-Dirac distribution (RC-FDD) (f). (g,h) {Peak height (g) and integrated peak intensity (h) of Q2 as a function of temperature. Here the RC-FDD divided spectra (f) were used. The integration was done from -0.125 eV to 0.005 eV.} (i) Same as (h), but plotted with the $\log(T)$ scale.
}
\label {Fig.S5}
\end{figure}

\begin{figure}[ht]
\includegraphics[width=.6\columnwidth]{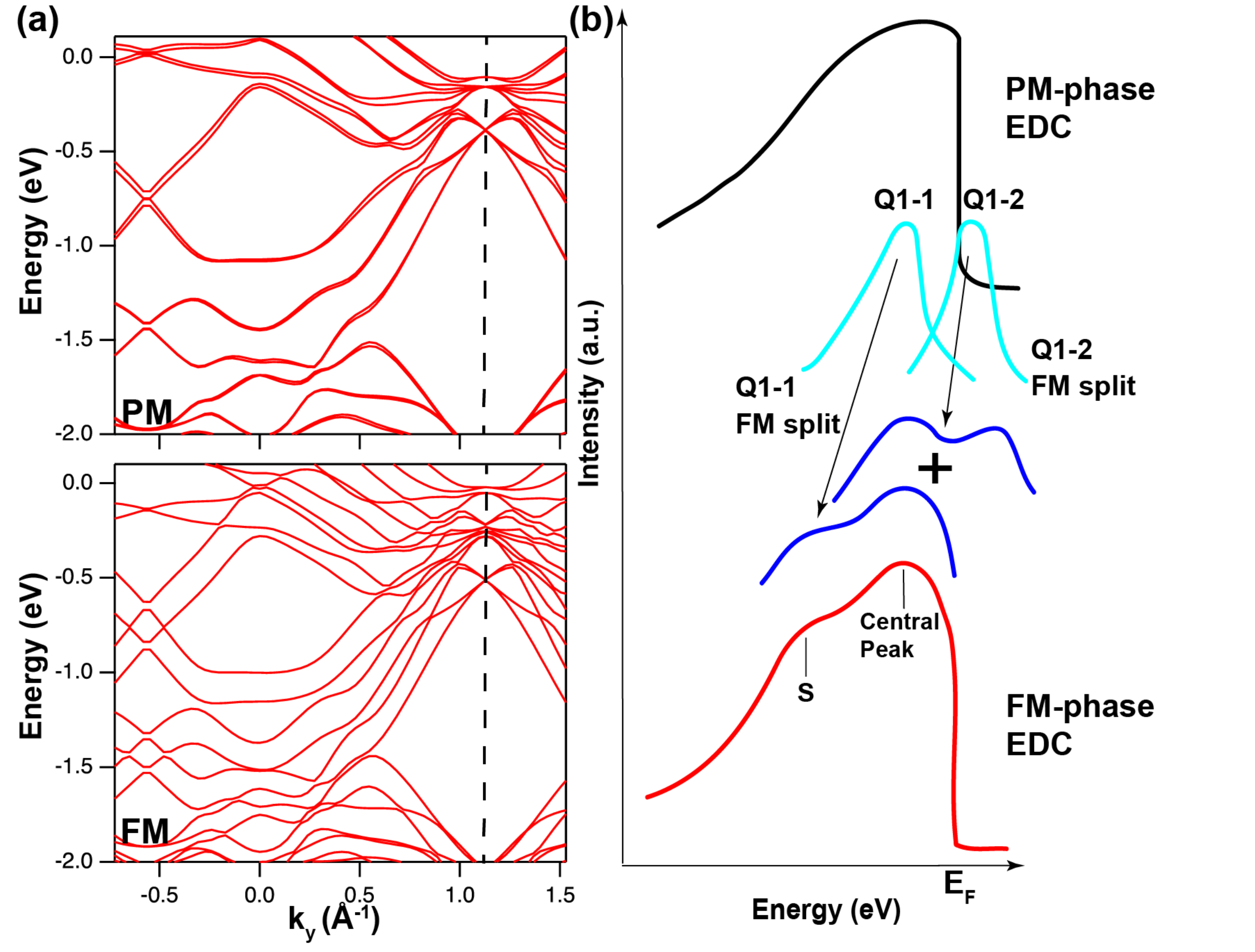}
\caption{ {(a) Calculated bands from DFT for the PM and FM phases (same as Fig. S3). The vertical dashed line indicates the momentum point corresponding to the experimental EDCs shown in Fig. 3(b). (b) A schematic showing how the FM exchange splitting of the Q1 peaks could lead to the observed spectral change in Fig. 3(b). Q1-1 and Q1-2 are two peaks that are already present in the PM phase (see the vertical dashed line in (a)). FM exchange splittings of Q1-1 and Q1-2 lead to emergence of a new satellite band (labelled S in Fig. S5(a)) at $\sim$-0.3 eV (from Q1-1), a more pronounced central peak at $\sim$-0.15 eV (due to overlapping contributions from split Q1-1 and Q1-2 peaks) and reduced density of states (DOS) at $E_F$ (as the minority band of Q1-2 is pushed above $E_F$).}
}
\label {Fig.S6}
\end{figure}

\clearpage